\documentclass[prd,amsmath,
twocolumn,floatfix,amssymb, preprintnumbers, nofootinbib, superscriptaddress]{revtex4}

\usepackage{amsmath,amssymb,amsfonts}
\usepackage{graphicx}
\usepackage{hyperref}
\usepackage{slashed}
\usepackage{soul}
\usepackage{booktabs,tabulary}
\usepackage{mciteplus}
\usepackage{color}

\usepackage{bibentry}
\newcommand{\ignore}[1]{}
\newcommand{\nobibentry}[1]{{\let\nocite\ignore\bibentry{#1}}}

\makeatletter
\def\bibinfo@X@title#1,{\ignorespaces}
\makeatother 

\begin{document}

\title{Dispersive analysis of the $\gamma^{*}\gamma^{*} \to \pi \pi$ process}

\author{Igor Danilkin}\email{danilkin@uni-mainz.de}
\author{Oleksandra Deineka}
\author{Marc Vanderhaeghen}
\affiliation{Institut f\"ur Kernphysik \& PRISMA$^+$  Cluster of Excellence, Johannes Gutenberg Universit\"at,  D-55099 Mainz, Germany}

\date{\today}

\begin{abstract}
We present a dispersive analysis of the double-virtual photon-photon scattering to two pions up to 1.5 GeV. Through unitarity, this process is very sensitive to hadronic final-state interaction. For the $s$-wave, we use a coupled-channel $\pi\pi,\, K\bar{K}$ analysis which allows a simultaneous description of both $f_0(500)$ and $f_{0}(980)$ resonances. For higher energies, $f_2(1270)$ shows up as a dominant structure which we approximate by a single-channel $\pi\pi$ rescattering in the $d$-wave. In the dispersive approach, the latter requires taking into account $t$- and $u$-channel vector-meson exchange left-hand cuts which exhibit an anomalous-like behavior for large space-like virtualities. We show how to readily incorporate such behavior using a contour deformation. Besides, we devote special attention to kinematic constraints of helicity amplitudes and show their correlations explicitly.
\end{abstract}

\maketitle

\section{Introduction}\label{intro}
It is still an open question whether a current ultra-precise $(g-2)_\mu$ measurement can probe the physics beyond the Standard Model. The presently observed $3-4\,\sigma$ deviation between theory \cite{Jegerlehner:2017gek,Keshavarzi:2018mgv,*Davier:2017zfy,Danilkin:2019mhd} and experiment \cite{Tanabashi:2018oca} has a potential to become more significant once results from new measurements at both FERMILAB \cite{LeeRoberts:2011zz,*Grange:2015fou} and J-PARC \cite{Iinuma:2011zz} are available. On the other hand, the current theoretical error results entirely from hadronic contributions. The hadronic uncertainties mainly originate from the hadronic vacuum polarization (HVP) and the hadronic light-by-light (HLbL) processes. Forthcoming data from the high luminosity $e^+e^-$ colliders, in particular, from the BESIII and Belle-II Collaborations will further reduce the uncertainty in the HVP in the coming years to make it commensurate with the experimental precision on $(g-2)_\mu$. The remaining hadronic uncertainty results from HLbL, where, apart from the pseudo-scalar pole contribution, a further nontrivial contribution comes from the two-particle intermediate states such as $\pi\pi$, $\pi\eta$ and $K\bar{K}$. 

The rescattering of $\pi\pi$ and $\pi\eta$ is responsible for the contribution from $f_0(500)$, $f_0(980)$, $f_2(1270)$, and $a_0(980)$, which can be taken into account in a dispersive framework. 
Among them, only $f_2(1270)$ can be interpreted within the quark model as a state that does not originate from long-range interactions \cite{frautschi1963regge}. Given the fact that it is relatively narrow, its contribution to $(g-2)_\mu$ can be accounted for in two ways: by using a pole contribution as it is given in \cite{Pauk:2014rta} (updated in \cite{Danilkin:2016hnh} using recent data from the Belle Collaboration \cite{Masuda:2015yoh}) or through fully dispersive formalisms \cite{Colangelo:2017fiz,*Colangelo:2017qdm,*Colangelo:2014pva} and \cite{Pauk:2014rfa} with input from $\gamma^*\gamma^* \to \pi\pi$. The comparison will shed light on the effective resonance description of other resonances such as axial-vector contributions \cite{Jegerlehner:2017gek,Pauk:2014rta}. 

In this paper, we present an analysis of the double virtual photon fusion reaction with pions in the final state. Our approach relies on the modified Muskhelishvili-Omn\`es formalism, which proves to be efficient in the description of the real photon data \cite{GarciaMartin:2010cw}. Within the maximal analyticity assumption \cite{Mandelstam:1959bc}, all of the non-analytic behavior of the amplitude should be coming from the unitarity and crossing symmetry constraints. Therefore in order to write the dispersion-integral representation for the partial wave helicity amplitudes, one needs to make sure that they are free from kinematic constraints at thresholds or pseudo-thresholds. The critical step in finding these constraints is the decomposition of the amplitude into Lorentz structures and invariant amplitudes \cite{Bardeen:1969aw, *Tarrach:1975tu,*Drechsel:1997xv}. The latter are  expected to satisfy the Mandelstam dispersion-integral representation \cite{Mandelstam:1958xc}. Once a suitable set of Lorentz structures is found, the rest is straightforward. Our work is a continuation of a previous work where, for the first time, the single virtual case for the $d$-wave has been studied \cite{Danilkin:2018qfn}. In the double virtual photon case, there is an additional complication related to the anomalous threshold behavior, as was pointed out in \cite{Hoferichter:2019nlq}. We will show an alternative way of taking this contribution into account using an appropriate contour deformation.

\section{Formalism}
\subsection{Kinematic constraints}

The two-photon fusion reaction $\gamma^*\gamma^*\to \pi\pi$ is a subprocess of the unpolarized double tagged process $e^+(k_1)e^{-}(k_2)\to e^{+}(k_1')e^{+}(k_2') \pi(p_1)\pi(p_2)$ which is given (in Lorenz gauge) as
\begin{eqnarray}\label{ee->eepipi_1}
i\,{\cal M}&=&\frac{i\,e^2}{q_1^2 q_2^2}\,[\bar{\upsilon}(k_1)\,\gamma_\mu\,\upsilon(k_1')]\,[\bar{u}(k_2')\,\gamma_\nu\,u(k_2)]\,H^{\mu\nu}\,,\\
H^{\mu\nu}&=&  i\int d^4 x\,e^{-i\,q_1\cdot x} \langle \pi(p_1)\pi(p_2)|T(j^{\mu}_{em}(x)
\,j^{\nu}_{em}(0))|0\rangle\,, \nonumber
\end{eqnarray}
with $q_1\equiv k_1-k_1'$, where the momenta of leptons $k^\prime_1$ and $k^\prime_2$ are detected. This corresponds to the kinematical situation where photons with momenta $q_1$ and $q_2$ have finite space-like virtualities, $q_1^2=-Q_1^2$ and $q_2^2=-Q_2^2$. By contracting the hadronic tensor $H^{\mu\nu}$ with polarization vectors, one defines helicity amplitudes $H_{\lambda_1 \lambda_2}$ which can be further decomposed into partial waves
\begin{eqnarray}\label{p.w.expansion}
&&\epsilon_\mu(q_1,\lambda_1)\,\epsilon_\nu(q_2,\lambda_2)\,H^{\mu\nu}\equiv e^{i\phi(\lambda_1-\lambda_2)}H_{\lambda_1 \lambda_2}\\&&=e^{i\phi(\lambda_1-\lambda_2)} N \sum_{J\,\text{even}}(2J+1)\,h^{(J)}_{\lambda_1\lambda_2}(s)\,d_{\Lambda,0}^{(J)}(\theta)\,,\nonumber
\end{eqnarray}
where $\Lambda=\lambda_1-\lambda_2$, $d_{\Lambda,0}^{(J)}(\theta)$ is a Wigner rotation function and $\theta$ is the c.m. scattering angle. In Eq. (\ref{p.w.expansion}), $N=1$ for $\gamma^*\gamma^*\to\pi\pi$ and $N=1/\sqrt{2}$ for $\gamma^*\gamma^*\to K \bar K$, ensuring the same unitarity relations for the identical and non-identical particles in the case of $I=0$.

It is well known that partial wave (p.w.) amplitudes $h^{(J)}_{\lambda_1\lambda_2}$ may have kinematic singularities
or may obey kinematic constraints \cite{Gasparyan:2010xz, Lutz:2011xc,*Heo:2014cja}. Therefore, it is important to find a transformation to a new set of amplitudes which are more appropriate to use in partial-wave dispersion relations. The key step is to decompose the scattering amplitude into a complete set of invariant amplitudes \cite{Bardeen:1969aw, *Tarrach:1975tu,*Drechsel:1997xv} (see also \cite{Colangelo:2015ama})
\begin{equation}
H^{\mu\nu}=\sum_{i=1}^{5}F_i\,L^{\mu\nu}_i\,,
\end{equation}
where
\begin{align}
\label{Eq:Lorez_structures}
&L_1^{\mu\nu}=q_1^{\nu}\,q_2^{\mu}-(q_1, q_2)\,g^{\mu\nu},\\
&L_2^{\mu\nu}=(\Delta^2\,(q_1, q_2)-2\,(q_1,
\Delta)\,(q_2,
\Delta))\,g^{\mu\nu}-\Delta^2\,q_1^\nu\,q_2^\mu
\nonumber\\  
& -2\,(q_1, q_2)\,\Delta^{\mu}\,\Delta^{\nu}
+2\,(q_2,\Delta)\,q_1^{\nu}\,\Delta^{\mu}+2(q_1, \Delta)\,q_2^{\mu}\,\Delta^{\nu}\,,\nonumber\\
&L_3^{\mu\nu}=(t-u)\Big\{\left(Q_1^2\,\left(q_2,\Delta\right)-Q_2^2\,\left( q_1, \Delta\right)\right) \left(g^{\mu \nu }-\frac{q_1^{\nu}q_2^{\mu} }{\left(q_1, q_2\right)}\right)\nonumber\\
&-\left(\Delta ^{\nu }-\frac{\left(q_2, \Delta \right) q_1^{\nu }}{\left(q_1, q_2\right)}\right) \left(Q_1^2 q_2^{\mu}+q_1^{\mu } \left(q_1, q_2\right)\right)\nonumber\\
& +\left(\Delta ^{\mu }-\frac{\left(q_1 , \Delta  \right) q_2^{\mu }}{\left(q_1, q_2\right)}\right) \left(Q_2^2\,q_1^{\nu }+q_2^{\nu } \left(q_1, q_2\right)\right)\Big\}\,,
\nonumber\\
&L_4^{\mu\nu}=Q_1^2\,Q_2^2\,g^{\mu \nu}+Q_1^2\,q_2^{\mu }\, q_2^{\nu }+Q_2^2\,q_1^{\mu }\,q_1^{\nu }+q_1^{\mu }\,q_2^{\nu }\, \left(q_1, q_2\right)\,,\nonumber\\
&L_5^{\mu\nu}=\left(Q_1^2\,\Delta ^{\mu }+\left(q_1,\Delta\right) q_1^{\mu }\right) \left(Q_2^2\,\Delta ^{\nu }+\left(q_2,\Delta \right) q_2^{\nu }\right)\,,\nonumber
\end{align}
where $\Delta \equiv p_1-p_2$ and each $L_i^{\mu\nu}$ satisfies a gauge invariance constraint, i.e., $q_{1\mu}\,L_i^{\mu\nu}=q_{2\nu}\,L_i^{\mu\nu}=0$. The numbering of the Lorentz structures is chosen such that in the single virtual case only $L^{\mu\nu}_{1,2,3}$ contribute to the process \cite{Danilkin:2018qfn}, while in the real photon case, only $L^{\mu\nu}_{1,2}$ are relevant, which coincide with the tensor structures used in \cite{Danilkin:2012ua, Danilkin:2017lyn,*Deineka:2018nuh}. The invariant amplitudes $F_i$ are free from kinematic singularities or constraints and depend on the Mandelstam variables, which we choose as $s=(q_1+q_2)^2$, $t=(p_1-q_1)^2$, and $u=(p_1-q_2)^2$. The prefactor $(t-u)$ in front of the tensor $L_3^{\mu\nu}$ is chosen so as to make all five amplitudes $F_{i}$ even under pion and photon crossing symmetry $(t\leftrightarrow u)$ \cite{Moussallam:2013una, Colangelo:2015ama}. We note that the Born terms possess a double pole structure in the soft-photon limit, as a manifestation of Low's theorem \cite{Low:1958sn}. The kinematic constraints can be obtained by analyzing projected helicity amplitudes in terms of the quantities
\begin{equation}\label{AnJ}
A_n^J(s)=\frac{1}{(p\,q)^J}\int_{-1}^{1}\frac{dz}{2}P_J(z)\,F_n(s,t)\,,
\end{equation}
which are free of any singularities due to the properties of the Legendre polynomials \cite{Lutz:2011xc,*Heo:2014cja}. In Eq. (\ref{AnJ}), $q$ and $p$ are initial and final relative momenta in the c.m. frame. Due to specifics of our basis (\ref{Eq:Lorez_structures}), all of the results below are given for the Born subtracted p.w. amplitudes 
\begin{align}
\bar{h}_{\lambda_1\lambda_2}^{(J)}\equiv h^{(J)}_{\lambda_1\lambda_2}-h^{(J), Born}_{\lambda_1\lambda_2},
\end{align}
where for $s$-wave it holds that \cite{Colangelo:2017fiz,*Colangelo:2017qdm,*Colangelo:2014pva}
\begin{align}\label{BarierFactors_Swave}
&\bar h^{(0)}_{++}(s)\pm \bar h^{(0)}_{00}(s) \sim (s-s_{\text{kin}}^{(\mp)})\,,\\
& s_{\text{kin}}^{(\pm)} \equiv -\left(Q_1\pm Q_2\right)^2\,,\nonumber
\end{align}
with $Q_i \equiv \sqrt{Q_i^2}$ ($i=1,2$). Note that in the single virtual and real photon cases these constraints are required by the soft-photon theorem \cite{Low:1958sn} and have been implemented already in \cite{Moussallam:2013una,Hoferichter:2011wk,Morgan:1987gv,*Dai:2014zta,*Dai:2014lza}. The kinematically uncorrelated amplitudes for the $s$-wave can be obtained by dividing the left-hand side (lhs) of  Eq. (\ref{BarierFactors_Swave}) by its right-hand side (rhs)
\begin{align}\label{new_pw_amplitudes}
&\bar h^{(0)}_{i=1,2}(s)=\frac{\bar h^{(0)}_{++}(s)\pm \bar h^{(0)}_{00}(s)}{s-s_{\text{kin}}^{(\mp)} }\,.
\end{align}
In \cite{Danilkin:2018qfn} the kinematically unconstrained basis of the partial wave amplitudes were derived for the single virtual case. Below we extend this result for the double-virtual case for $J=2$,
\begin{align}\label{Eq.newampl}
&(s+Q_1^2+Q_2^2)\,\bar{h}_{+-}^{(2)}+2 \sqrt{2} Q_1^2 Q_2^2\,\bar{h}_-\sim \gamma_1(s) \,,\\
&\sqrt{2}\,\bar{h}_{+-}^{(2)}-\bar{h}_++\left(Q_1^2+Q_2^2\right)\,\bar{h}_-\sim \gamma_1(s)\,, \nonumber \\
&\sqrt{2}\,\bar{h}_{+-}^{(2)}+\left(s+Q_1^2+Q_2^2\right)\,\bar{h}_-\sim \gamma_1(s)\,,\nonumber\\
&\sqrt{6}\,s\,\bar{h}_{+-}^{(2)}-2 \sqrt{3}\,s\,\bar{h}_{+}+3\,s\,(s+Q_1^2+Q_2^2)\,\bar{h}_0+6\,s\,\bar{h}_{++}^{(2)} \nonumber\\
& \hspace{0.5cm}+\sqrt{3} \left(s^2+2 \left(Q_1^2+Q_2^2\right) s-\left(Q_1^2-Q_2^2\right)^2\right)\,\bar{h}_{-} \sim \gamma_2(s)\,,\nonumber\\
&6s\,(s+Q_1^2+Q_2^2)\,\bar{h}_{++}^{(2)}+12\,Q_1^2\,Q_2^2\,s\, \bar{h}_{0}\nonumber\\
& \hspace{1.7cm}-\sqrt{6}\left(s\,(Q_1^2+Q_2^2)+(Q_1^2-Q_2^2)^2\right)\,\bar{h}_{+-}^{(2)}\nonumber \\
& \hspace{1.7cm}+2\sqrt{3} \left(s\,(Q_1^2+Q_2^2)+(Q_1^2-Q_2^2)^2\right)\,\bar{h}_{+}\nonumber\\
& \hspace{1.7cm}-2\sqrt{3} \left(Q_1^2-Q_2^2\right)^2  (s+Q_1^2+Q_2^2)\,\bar{h}_{-} \sim \gamma_2(s)\,,\nonumber
\end{align}
with
\begin{align}
\gamma_n(s)\equiv \lambda^n(s,-Q_1^2,-Q_2^2)\,(s-4\,m_\pi^2)\,,
\end{align}
where $\lambda$ is the K\"all\'en triangle function and  $\bar{h}_{+,-,0}$ were introduced for convenience 
\begin{align}
&{\bar{h}}_{+}(s)\equiv\frac{\sqrt{s}}{Q_2}\,\bar{h}^{(2)}_{+0}(s)+\frac{\sqrt{s}}{Q_1}\,{\bar{h}}^{(2)}_{0+}(s), \\
&{\bar{h}}_{-}(s)\equiv \left(\frac{\sqrt{s}}{Q_2}\,{\bar{h}}^{(2)}_{+0}(s)-\frac{\sqrt{s}}{Q_1}\,{\bar{h}}^{(2)}_{0+}(s)\right)\frac{1}{Q_1^2-Q_2^2}\,,\nonumber\\
&{\bar{h}}_{0}(s)\equiv\frac{{\bar{h}}^{(2)}_{00}(s)}{Q_1\,Q_2}\,. \nonumber
\end{align}
We emphasize, that in addition to the $s_{\text{kin}}^{(\pm)}$ points, the  p.w. amplitudes for $J \neq 0$ exhibit a so-called centrifugal barrier factor at $4\,m_\pi^2$. The new set of amplitudes $\bar{h}^{(2)}_{i=1..5}(s)$ we obtain as in (\ref{new_pw_amplitudes}) by dividing the lhs of Eq. (\ref{Eq.newampl}) by its rhs.\footnote{
Note that when $Q_1^2 = Q_2^2$ (and pions are in the final state), special care is required. In that case, $H_{+0}=-H_{0+}$, and only four Lorentz tensors in Eq. (\ref{Eq:Lorez_structures}) are independent. Therefore, one needs to reshuffle Eq. (\ref{Eq.newampl}) in such a way that only four amplitudes $\bar{h}^{(J)}_i$ survive. We checked that numerically the results for $Q_1^2\approx Q_2^2$ given by Eq. (\ref{Eq.newampl}) are consistent with the strict $Q_1^2 = Q_2^2$ limit.} We emphasize that Eq. (\ref{Eq.newampl}) shows the correlation of the p.w. helicity amplitudes explicitly, as compared with the result based on the Roy-Steiner equations \cite{Colangelo:2014dfa, Hoferichter:2019nlq}, where kinematic constraints are contained in the integral kernels. The full set of these off-diagonal kernels is given in \cite{Hoferichter:2019nlq}, and the final solution is obtained by diagonalization of the kernel matrix.

\subsection{Dispersion relations}

The new set of amplitudes $\bar{h}^{(J)}_{1-5}$ contains only dynamical singularities. These are right and left-hand cuts, and one can write a dispersion relation in the following form (modulo subtractions which will be discussed in Section \ref{Section:results})
\begin{align}\label{DR_1}
\bar{h}^{(J)}_{i}(s)
&=\int_{-\infty}^{0}\frac{d s'}{\pi}\frac{\text{Disc}\,\bar{h}^{(J)}_{i}(s')}{s'-s}+\int_{4m_\pi^2}^{\infty} \frac{ds'}{\pi}\frac{\text{Disc}\,h^{(J)}_{i}(s')}{s'-s}\,,
\end{align}
where we noted that $\text{Disc}\,\bar{h}^{(J)}_{i}(s)=\text{Disc}\,h_i^{(J)}(s)$ along the right-hand cut. The latter is determined by the unitarity condition and in the elastic approximation is given by
\begin{align}\label{h_Unitarity}
&\text{Disc}\,h^{(J)}_{i}(s)=t^{(J)*}(s)\,\rho(s)\,h^{(J)}_{i}(s)\,,\\
&\rho(s)=\frac{p(s)}{8\,\pi\sqrt{s}}\,\theta(s-4\,m_\pi^2)\,,\nonumber
\end{align}
where $\rho(s)$ is a two-body phase space factor and $t^{(J)}(s)$ is the hadronic scattering amplitude, which is normalized as $\text{Im}(t^{(J)})^{-1}=-\rho$. 
For the energy region above 1 GeV, it is necessary to take into account the inelasticity. The first relevant inelastic channel is $K\bar{K}$, which is required to capture the dynamics of the $f_0(980)$ scalar meson. For the coupled-channel case, the phase-space function $\rho(s)$ and the amplitude $t^{(J)}(s)$ turn into $(2\times 2)$ matrices, while $h^{(J)}_{i}$ will be written in the $(2\times 1)$ form with elements $h^{(J)}_{i}$ and $k^{(J)}_{i}$ which correspond to the $\gamma^*\gamma^*\to \pi\pi$ and $\gamma^*\gamma^* \to K\bar{K}$ amplitudes, respectively. The solution to Eq. (\ref{DR_1}) is given by the well known Muskhelishvili-Omn\`es method for treating the final-state interactions \cite{Omnes:1958hv}. It is based on writing a dispersion relation for $\bar{h}^{(J)}_{i}(\Omega^{(J)})^{-1}$ \cite{GarciaMartin:2010cw}, where $\Omega^{(J)}$ is the Omn\`es function which satisfies a similar unitarity constraint,
\begin{equation}\label{Omnes_CC_Unitarity}
\text{Disc}\,\Omega^{(J)}(s)=t^{(J)}(s)\,\rho(s)\,\Omega^{(J)*}(s)\,.
\end{equation} 
As a result, we obtain
\begin{align}\label{Rescattring:general}
&{h}^{(J)}_{i}(s)= {h}^{(J),\text{Born}}_{i}(s)
\\ 
&\hspace{0.6cm}+\Omega^{(J)}(s) \bigg[
-\int_{4m_\pi^2}^{\infty}\frac{ds'}{\pi}\,\frac{\text{Disc}\,(\Omega^{(J)}(s'))^{-1}\,{h}^{(J),\text{Born}}_{i}(s')}{s'-s}
\nonumber \\
& \hspace{0.6cm}+
\int_{-\infty}^{0}
\frac{ds'}{\pi}\,\frac{(\Omega^{(J)}(s'))^{-1}\,\text{Disc}\,{\bar{h}}^{(J)}_{i}(s')}{s'-s}
\bigg]\,, \nonumber 
\end{align}
which can be straightforwardly generalized for the coupled-channel case. The Born subtracted amplitudes along the left-hand cut (the second term inside the brackets) are given by multi-pion exchanges in the $t$ and $u$ channels, which in practice can be approximated by resonance $(R)$ exchanges \cite{GarciaMartin:2010cw}. The dominant contribution is generated by vector mesons $\omega$ and $\rho$. The contribution from other heavier resonances will be absorbed in an effective way by allowing for a slight adjustment of the $VP\gamma$ coupling \cite{Danilkin:2018qfn}. 

Here we note that there is freedom in writing the dispersion relation. In principle, one could write a dispersion relation for the combination $(\bar{h}^{(J)}_{i}-h^{(J),V}_{i})(\Omega^{(J)})^{-1}$, as was done for $\gamma\gamma^* \to \pi\pi$ in \cite{Moussallam:2013una}. However, in this case, one needs to make an assumption about the high-energy dependence of the real part of $h^{(J),V}_{i}$, as was explained in \cite{Hoferichter:2019nlq}. In this work, we take out only the Born term in Eg. (\ref{Rescattring:general}) and therefore need to know only the high energy behavior of the imaginary part of the vector mesons exchange entering the left-hand cut, which does not have any polynomial ambiguity \cite{GarciaMartin:2010cw}.

\subsection{Left-hand cuts} 

The generalization of the Born contribution to the case of off-shell photons is performed by multiplying the scalar QED result by the electromagnetic pion (kaon) form factors \cite{Fearing:1996gs,Colangelo:2015ama} which lead to the following invariant amplitudes
\begin{align}\label{Fi:Born}
&F_1^{\text{Born}}=-\frac{e^2 \left(4\,m_i^2+Q_1^2+Q_2^2\right)}{\left(t-m_i^2\right) \left(u-m_i^2\right)}\,f_{i}(Q_1^2)\,f_{i}(Q_2^2)\,,\quad \\
&F_2^{\text{Born}}=-\frac{e^2}{\left(t-m_i^2\right) \left(u-m_i^2\right)}\,f_{i}(Q_1^2)\,f_{i}(Q_2^2)\,,
\nonumber \\
&F_3^{\text{Born}}=F_4^{\text{Born}}=F_5^{\text{Born}}=0\,,\nonumber
\end{align}
where $i=\pi\,(K)$ for
$\gamma^*\gamma^*\to \pi \pi$ ($K \bar K$). As these Born terms coincide with the pion pole terms obtained in a dispersive derivation, there is full agreement between the results of \cite{Fearing:1996gs,Colangelo:2015ama}. We note that the double pole structure of the Born amplitudes does not bring extra complications to Eq.~(\ref{Rescattring:general}) since its singularities lie outside the physical region. The electromagnetic spacelike pion and kaon form factors in the region $Q^2 \lesssim 1$ GeV$^2$ are parameterized by simple monopole forms yielding the following mass parameters: $\Lambda_{\pi}=0.727(5)$ GeV and $\Lambda_K=0.872(47)$ GeV with $\chi^2/\text{d.o.f.}=1.22$ \cite{Dally:1981ur, *Amendolia:1984nz, *Tadevosyan:2007yd} and $\chi^2/\text{d.o.f.}=0.69$\cite{Dally:1980dj, *Amendolia:1986ui, *Carmignotto:2018uqj}, respectively. 

The vector-meson exchange left-hand cuts are obtained by the effective Lagrangian which couples photon, vector ($V$), and pseudoscalar ($P$) meson fields,
\begin{equation}\label{LVPg}
{\cal  L}_{VP\gamma}=e\,C_{VP\gamma}\,\epsilon^{\mu\nu\alpha\beta}\,F_{\mu\nu}\,\partial_\alpha P\,V_\beta\,,
\end{equation}
where $F_{\mu\nu}=\partial_\mu\,A_\nu-\partial_\nu\,A_\mu$. This Lagrangian density implies 
\begin{align}\label{Fi:Vexch}
&F_1^{V \text{exch}}=-\sum _V \frac{e^2\,C_{VP\gamma}^2}{2} \left(\frac{4\,t+Q_1^2+Q_2^2}{t-m_V^2}\right. 
 \\
&\hspace{3cm}\left.+\frac{4\,u+Q_1^2+Q_2^2}{u-m_V^2}\right)\tilde{f}_{V,i}(Q_1^2,Q_2^2)\,,\quad \nonumber \\
&F_2^{V \text{exch}}=\sum _V \frac{e^2\,C_{VP\gamma}^2}{2} \left(\frac{1}{t-m_V^2}+\frac{1}{u-m_V^2}\right)\tilde{f}_{V,i}(Q_1^2,Q_2^2)\,,\nonumber \\
&F_3^{V \text{exch}}=\sum _V \frac{e^2\,C_{VP\gamma}^2}{t-u}\left(\frac{1}{u-m_V^2}-\frac{1}{t-m_V^2}\right)\tilde{f}_{V,i}(Q_1^2,Q_2^2)\,,\nonumber \\
&F_4^{V \text{exch}}=\sum _V e^2\,C_{VP\gamma}^2\, \left(\frac{1}{t-m_V^2}+\frac{1}{u-m_V^2}\right)
\tilde{f}_{V,i}(Q_1^2,Q_2^2)\,, \nonumber \\
&F_5^{V \text{exch}}=0\,, \nonumber \\
&\tilde{f}_{V,i}(Q_1^2,Q_2^2)\equiv f_{V,i}(Q_1^2)\,f_{V,i}(Q_2^2)\,, \nonumber 
\end{align}
where in the following we will use $g_{VP\gamma}\simeq C_{\rho^{\pm,0}\pi^{\pm,0}\gamma}\simeq C_{\omega\pi^{0}\gamma}/3 $ as the only fit parameter, as discussed in \cite{Danilkin:2018qfn}, yielding $g_{VP\gamma} =0.33$ GeV$^{-1}$. This value lies within $10\%$ with the Particle Data Group (PDG) average $g^{\text{PDG}}_{VP\gamma}=0.37(2)$ \cite{Tanabashi:2018oca}, thus justifying the approximation of left-hand cuts by vector mesons. The slight difference accounts for the contribution from other heavier left-hand cuts, which in general should be taken into account by imposing Regge asymptotics. Such a study is, however, beyond the scope of this analysis. In  Eq.~(\ref{Fi:Vexch}) $f_{V,\pi}(Q_i^2)$ are vector meson transition form factors.  For the $\omega$, we use the dispersive analysis from \cite{Danilkin:2014cra} (see also \cite{Schneider:2012ez}), while for the $\rho$ (sub-dominant) contribution we use the vector meson dominance model \cite{Sakurai1969}. We note, that the form factors are well defined only for the pole contribution. Using the fixed-$s$ Mandelstam representation, one can show that the vector pole contribution corresponds to replacing $t$ and $u$ with $m_V^2$ in the numerators of Eq.~(\ref{Fi:Vexch}). This is different compared to  Eq.~(\ref{Fi:Born}), where the pion pole contribution coincides exactly with the Born contribution as discussed above. We emphasize that for the dispersion relations written in the form of Eq. (\ref{Rescattring:general}) only $\text{Disc}\, h^{(J),V}_{\lambda_1\lambda_2}(s)$ is required as input, which is unique for the vector-pole contribution.

\subsection{Analytic structure of the left-hand cuts}

\begin{figure}[t]
\centering
\includegraphics[width =0.45\textwidth]{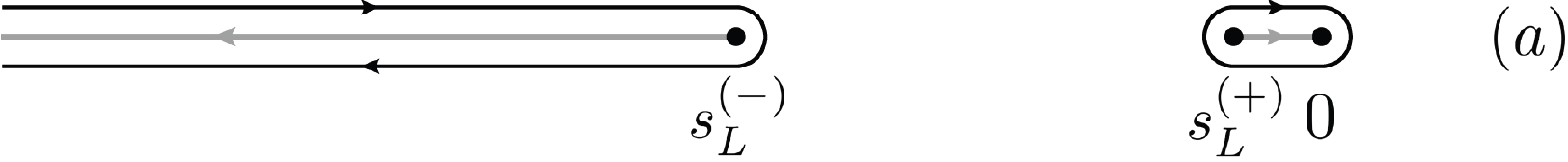}
\vspace{2mm}
\includegraphics[width =0.45\textwidth]{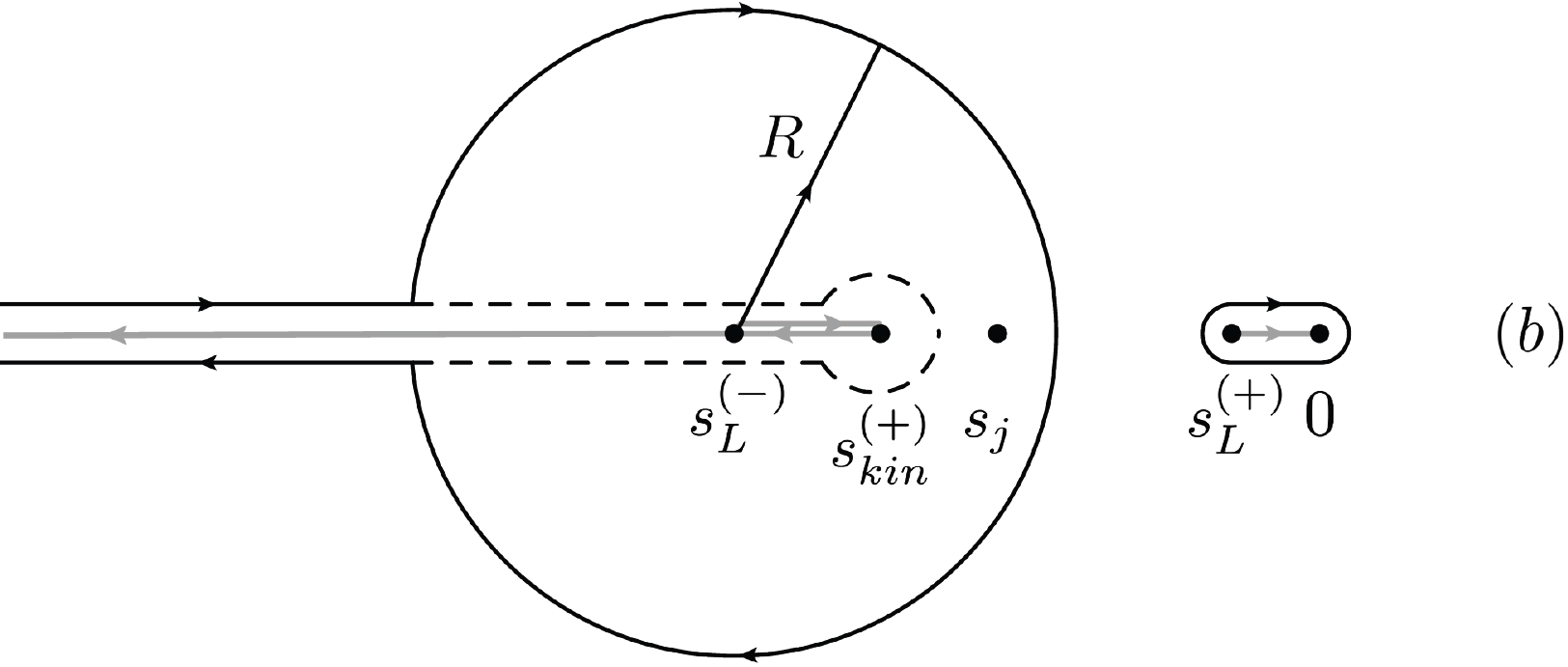}
\caption{Left-hand cut singularities and the integration contour for non-anomaly case (a) and its deformation for the anomaly case (b). See text for details.}
\label{Fig:CutStructure}
\end{figure}

In order to find a solution of the dispersion relations given in Eq. (\ref{Rescattring:general}), one needs to understand the singularity structure of the p.w. amplitudes $h_i^{(J)}$ as a function of the complex variable $s$.  For space-like photons, the p.w. Born amplitudes are real functions above the threshold and do not bring any complexity. On the other hand, the vector-meson exchange left-hand cut is determined by four branching points: $s=0$, $s=-\infty$, and
\begin{align}\label{s_L}
s_L^{(\pm)}&=\frac{1}{2}\big(2m_\pi^2-Q_1^2-Q_2^2-m_V^2-\frac{(m_\pi^2+Q_1^2)(m_\pi^2+Q_2^2)}{m_V^2}\big)\nonumber\\
&\pm \frac{\lambda^{1/2}(m_V^2,m_\pi^2,-Q_1^2)\,\lambda^{1/2}(m_V^2,m_\pi^2,-Q_2^2)}{2\,m_V^2}\,.
\end{align}
When one photon is real, the cut consists of two pieces: $(-\infty,s_L^{(-)}]$ and $[s_L^{(+)},0]$. However, when both photons carry a space-like virtuality, one has to be careful since for $Q_1^2\,Q_2^2 > (m_V^2-m_\pi^2)^2$ the left-hand branch point $s_L^{(-)}$ moves to the right and reaches the pseudo-threshold point $s_{kin}^{(+)}$ and only then moves to the left (see Fig.\ref{Fig:CutStructure}). In this case, the integration along the cut acquires an additional piece $[s_L^{(-)},\,s_{\text{kin}}^{(+)}]$, which is related to an "anomalous" discontinuity \cite{Karplus:1958zz,*Mandelstam:1960zz}. In addition, the integral around $s_{\text{kin}}^{(+)}$, in general, is nonzero and requires special care \cite{Hoferichter:2019nlq}. Indeed, according to Eq. (\ref{Eq.newampl}), the $J=2$ p.w. amplitude, that schematically is
\begin{align}\label{left-hand_test}
&h^V(s)=\frac{1}{(s-s_{\text{kin}}^{(+)})^2} \int_{-1}^{1}\frac{z^4\,d z}{t(s,z)-m_V^2}\,,
\end{align}
behaves like $\left(s-s_{\text{kin}}^{(+)}\right)^{-9/2}$. Splitting the contour path into an integral up to $s_{\text{kin}}^{(+)}-\epsilon$ and a circular integral of radius $\epsilon$ around $s_{\text{kin}}^{(+)}$ (dashed curve in Fig.\ref{Fig:CutStructure}) produces the cancellation of two singular pieces. In \cite{Hoferichter:2019nlq}, this was solved by using a fit function (which consists of an appropriate square-root-like behavior and a polynomial) in the vicinity of the singular point. We follow here a different strategy and enlarge the contour around $s_{\text{kin}}^{(+)}$ such that one stays away from possible numerical issues related to the anomaly piece (see Fig.\ref{Fig:CutStructure}). We propose to present $h^V(s)$ in the physical region as
\begin{align}\label{withanom}
h^V(s)&=\int_{-\infty}^{s_L^{(-)}-R}\frac{d\,s'}{\pi}\frac{\text{Disc}\,h^V(s')}{s'-s}+\int_{C_R}\frac{d\,s'}{2\pi i}\frac{h^V(s')}{s'-s}\nonumber\\
&+\int_{s_L^{(+)}}^{0}\frac{d\,s'}{\pi}\frac{\text{Disc}\,h^V(s')}{s'-s}
\end{align}
where $R$ is chosen such that $s_j=-Q_1^2 -Q_2^2 + 2\,m_\pi^2  - 2\,m_V^2$ lies inside the circle. The location of $s_j$ is determined by the condition that the imaginary part of the logarithm in Eq. (\ref{left-hand_test}) changes sign and therefore requires a proper choice of the Riemann sheet which we want to avoid. The merit of Eq. (\ref{withanom}) is such that it works for both anomaly and non-anomaly cases, so one can use it for any space-like $Q_i$ including the "transition" line when $Q_1^2\,Q_2^2=(m_V^2-m_\pi^2)^2$. In addition, it is independent on the degree of singularity and can be used equally well for higher p.w. with $J>2$. The generalization to the physical case with Omn\`es functions (\ref{Rescattring:general}) is then straightforward since all of the quantities are well defined at complex energies. 

For time-like virtualities (which are not of interest in this work) we refer the reader to \cite{Moussallam:2013una, Hoferichter:2013ama}, where different cases of overlapping left- and right-hand cuts are considered.

\subsection{Hadronic input}
For the $s$-wave isospin $I=0$ $(I=2)$ amplitude we use the coupled-channel (single channel) Omn\`es function from a dispersive summation scheme \cite{Gasparyan:2010xz, Danilkin:2010xd} which implements constraints from analyticity and unitarity. The method is based on the $N/D$ ansatz \cite{Chew:1960iv}, where the set of coupled-channel (single channel) integral equations for the $N$-function are solved numerically with the input from the left-hand cuts which we present in a model-independent form as an expansion in a suitably constructed conformal mapping variable. These coefficients, in principle, can be matched to $\chi$PT at low energy \cite{Danilkin:2011fz,*Danilkin:2012ap}. Here we use a data-driven approach and determine these coefficients directly from fitting to Roy analyses for $\pi\pi \to \pi\pi$ \cite{GarciaMartin:2011cn}, $\pi\pi \to K\bar{K}$ (for $I=0$) \cite{Buettiker:2003pp,*Pelaez:2018qny}, and existing experimental data for these channels. After solving the linear integral equation for $N(s)$, the $D$-function (the inverse of the Omn\`es function) is computed; more details will be given elsewhere \cite{Danilkin:2019}. 

For the $d$-wave $I=0,2$ amplitudes, we use the single-channel Omn\`es function in terms of the corresponding phase shifts,
\begin{equation}\label{OmenesPhaseShift}
\Omega_I^{(2)}(s)=\exp\left(\frac{s}{\pi}\int_{4m_\pi^2}^{\infty} \frac{d s'}{s'}\frac{\delta_{I}^{(2)}(s')}{s'-s}\right).
\end{equation}
Its numerical evaluation requires a high-energy parametrization of the phase shifts. We use a recent Roy analysis \cite{GarciaMartin:2011cn} below 1.42 GeV, and let the phase smoothly approach $\pi$ ($0$) for $I=0$ ($I=2$) respectively.

\section{Discussion and results}
\label{Section:results}

\begin{figure}[!t]
\centering
\includegraphics[width =0.475\textwidth]{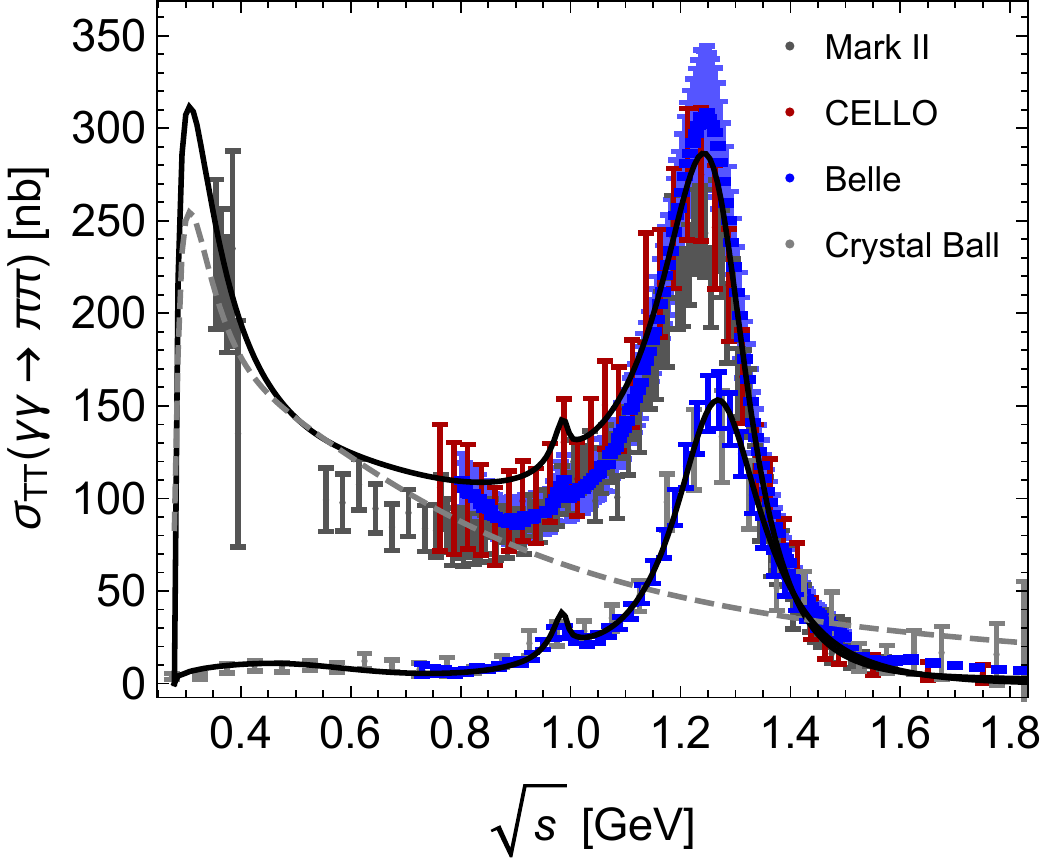}
\caption{Total cross sections for $\gamma\gamma \to \pi^+\pi^-$ ($|\cos\theta|<0.6$) (upper curve) and $\gamma\gamma \to \pi^0\pi^0$ ($|\cos\theta|<0.8$) (lower curve). The Born result is shown as dashed gray curves. 
The data are taken from
\cite{Mori:2007bu,*Uehara:2009cka,*Boyer:1990vu,*Behrend:1992hy,*Marsiske:1990hx}.
\label{fig:Q2=0}}
\end{figure}

\begin{figure*}[!t]
\centering
\includegraphics[width =0.45\textwidth]{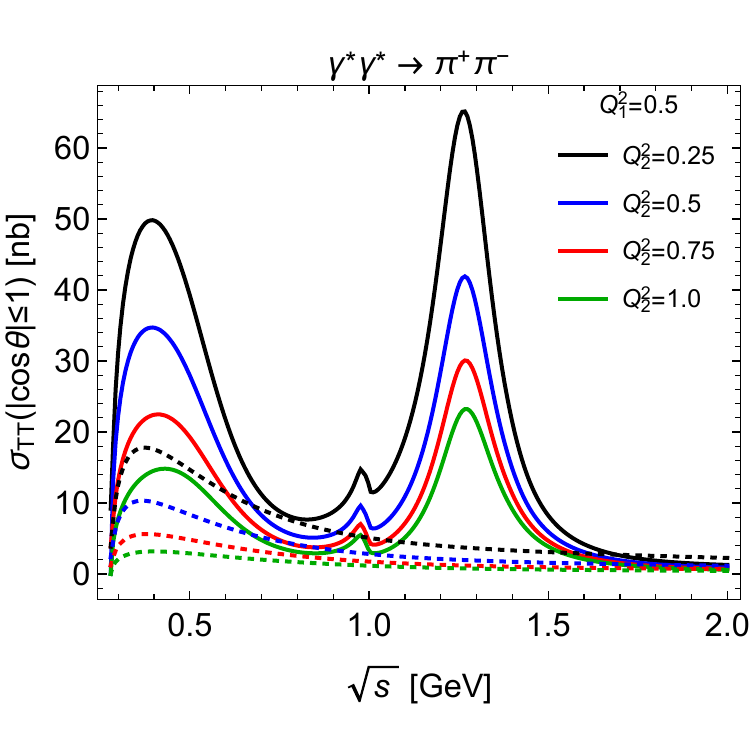}\quad 
\includegraphics[width =0.45\textwidth]{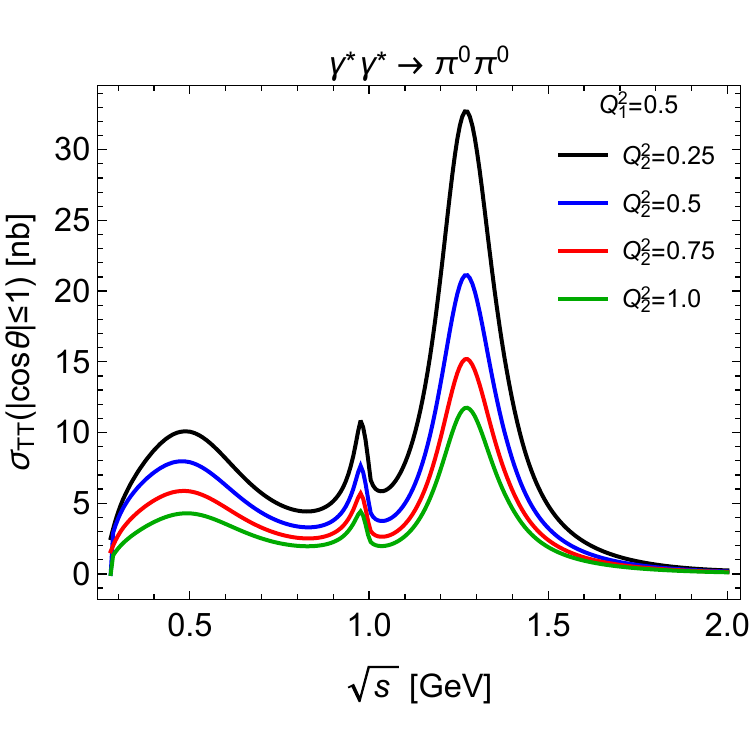}\\
\includegraphics[width =0.45\textwidth]{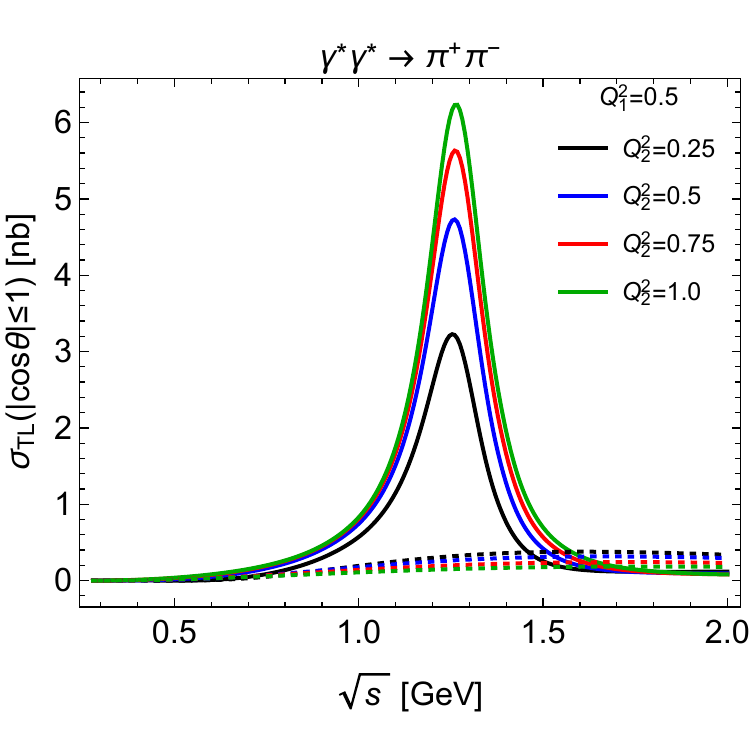}\quad 
\includegraphics[width =0.45\textwidth]{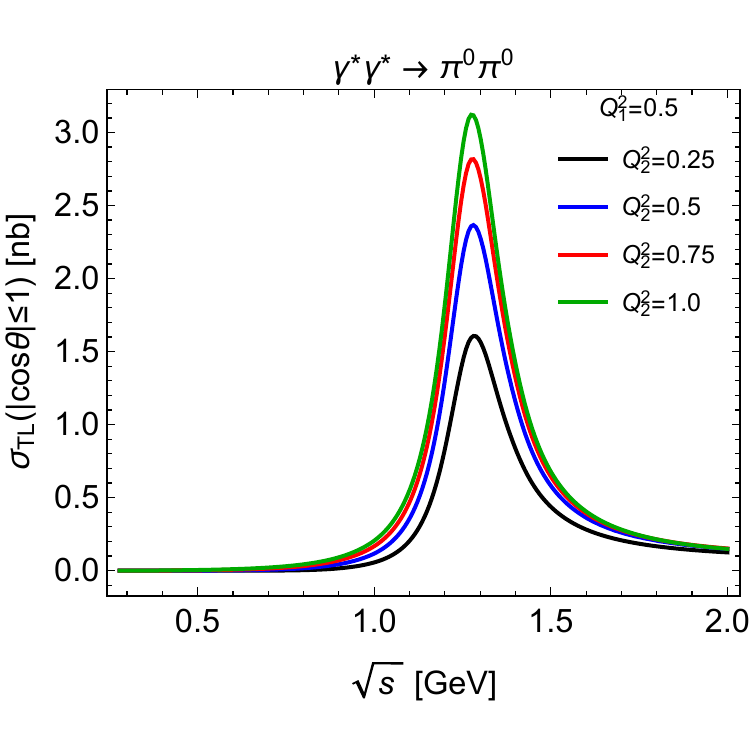}\\
\includegraphics[width =0.45\textwidth]{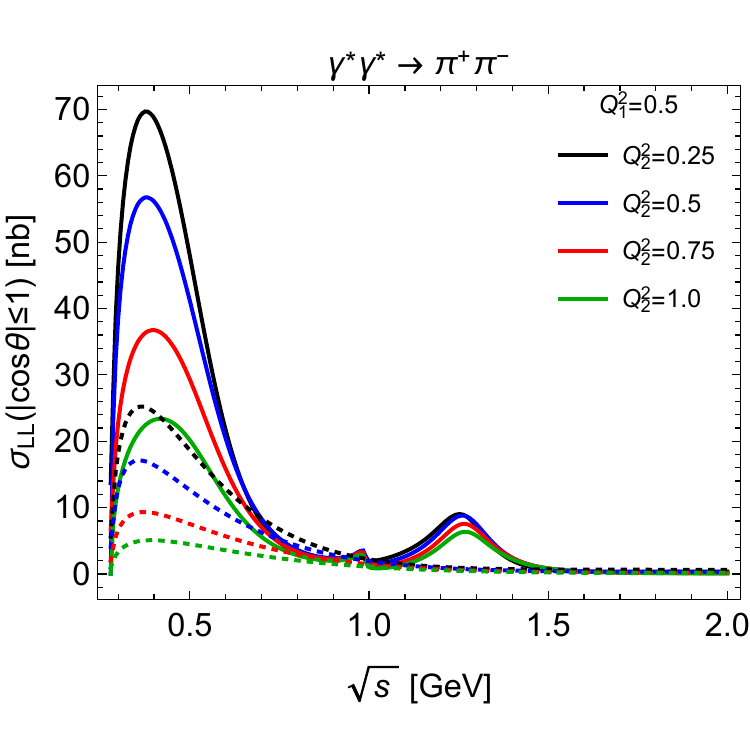}\quad 
\includegraphics[width =0.45\textwidth]{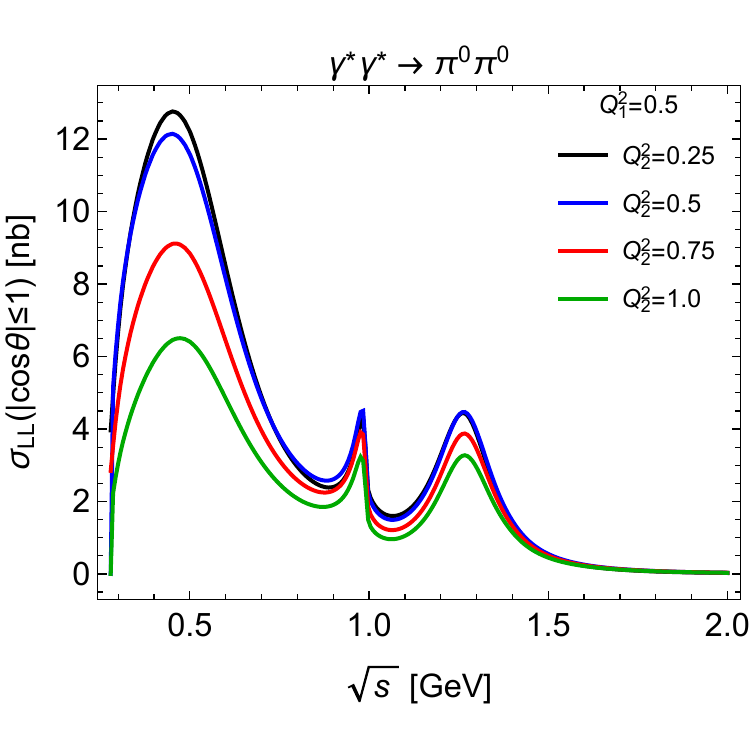}
\caption{Predictions for $\sigma_{TT}$, $\sigma_{TL}$, $\sigma_{LL}$ cross sections for $\gamma^*\gamma^* \to \pi^+\pi^-$ (left panels) and $\gamma^*\gamma^* \to \pi^0\pi^0$ (right panels) for $Q_1^2=0.5$ GeV$^2$ and $Q_2^2=0.25,0.5,0.75,1.0$ GeV$^2$ and for full angular coverage $|\cos\theta| \leq 1$. The Born results are shown as dotted curves.\label{fig:pipiQ1Q2}}
\end{figure*}

In Figs. \ref{fig:Q2=0} and \ref{fig:pipiQ1Q2}, we plot the $\gamma^*\gamma^* \to \pi\pi$ cross sections which involve either two transverse ($TT$) photon polarizations or two longitudinal ($LL$)  photon polarizations or one transverse and one longitudinal ($TL$) photon polarization defined by
\begin{align}\label{Eq:Cross_section}
&\frac{d \sigma_{TT}}{d \cos\theta}=\frac{\beta_{\pi\pi}}{64\,\pi\,\lambda^{1/2}(s,-Q_1^2,-Q_2^2)}\left(|H_{++}|^2+|H_{+-}|^2\right)\,,\nonumber\\
&\frac{d \sigma_{TL}}{d \cos\theta}=\frac{\beta_{\pi\pi}}{32\,\pi\,\lambda^{1/2}(s,-Q_1^2,-Q_2^2)}\, |H_{+0}|^2\,,\\
&\frac{d \sigma_{LL}}{d \cos\theta}=\frac{\beta_{\pi\pi}}{32\,\pi\,\lambda^{1/2}(s,-Q_1^2,-Q_2^2)}\, |H_{00}|^2\,,\nonumber\\ 
&\beta_{\pi\pi}=\frac{2\,p}{\sqrt{s}} \nonumber
\end{align}
where for the the neutral pions one has to include a symmetry factor of 1/2. The quantities $\sigma_{TT}$, $\sigma_{TL}$, $\sigma_{LT}$ and $\sigma_{LL}$ enter the cross section for the process $e^+e^- \to e^+e^- \pi\pi$ given in Refs. \cite{Budnev:1974de,Pascalutsa:2012pr}. It sets the convention for the flux factor, while the convention for the wave functions of the longitudinally polarized photons is chosen as
\begin{align}
&\epsilon^\mu(q_1,0) = \frac{1}{Q_1}\,\left(q, 0, 0, E_{q_1}\right)\,,\\
& \epsilon^\nu(q_2,0) = \frac{1}{Q_2}\,\left(-q, 0, 0, E_{q_2}\right)\,,\nonumber\\
&E_{q_i}=\sqrt{q^2-Q_i^2}\,,\quad  q=\frac{\lambda^{1/2}(s,-Q_1^2,-Q_2^2)}{2\sqrt{s}}\,. \nonumber
\nonumber
\end{align}
This convention reproduces continuously the real photon limit. 

Using unsubtracted dispersion relations, we postdict the cross-sections for the real photon case and 
give predictions for finite virtualities. We implement rescattering in  $s$- and $d$-waves, while the partial waves beyond are approximated by the Born terms. Including Born left-hand cuts alone predicts a reasonable description of the $f_0(500)$ and $f_0(980)$ regions; however, it fails to describe the $f_2(1270)$ resonance. For the latter, the inclusion of heavier left-hands cuts is necessary \cite{GarciaMartin:2010cw}. Following our previous work \cite{Danilkin:2018qfn}, we approximate them with only vector mesons exchanges and slightly adjust the coupling $g_{VP\gamma}$ in Eq.(\ref{Fi:Vexch}) to reproduce the $f_2(1270)$ peak in the $\gamma\gamma\to \pi^0\pi^0$ cross-section. We emphasize that this is the only parameter that we adjust to the real photon data to get a nice overall agreement (see Fig.\ref{fig:Q2=0}). We also note that the convergence of the unsubtracted dispersive integrals for $J=2$ is, in general, better than for $J=0$ due to the centrifugal barrier factor. Therefore, including vector meson left-hand cuts in the $s$-wave requires adding at least one subtraction, which can be fixed from chiral perturbation theory ($\chi$PT). We checked that for relatively small $Q^2$, the results of the two solutions are very similar. Since the finite $Q^2$ prediction from $\chi$PT is expected to show large corrections for $Q^2>0.25$ GeV$^2$, we decided to stay with the unsubtracted dispersion relation. In the present letter we show a selected result\footnote{The preliminary plots for $Q_1^2=Q_2^2=0.5$ GeV$^2$ shown in~\cite{Danilkin:2019mhd} suffered from a numerical instability in the calculation of one of the five dispersive integrals, which led to an overestimation of $\sigma_{LL}$ in the $f_2(1270)$ region, leaving the predictions for $\sigma_{TT}$ and $\sigma_{TL}$ mainly unchanged.} for a fixed value $Q_1^2=0.5$ GeV$^2$ for one photon virtuality and different values $Q_2^2=0.25, 0.5, 0.75, 1.0$ GeV$^2$ for the second photon virtuality (see Fig.\ref{fig:pipiQ1Q2}). The last two $Q_2^2$ points are above the anomaly point. For $\sigma_{TT}$ and $\sigma_{LL}$, we emphasize the importance of unitarization, which significantly increases the pure Born prediction at low energy. For $\sigma_{TL}$, we notice that the  helicity-1 contribution increases with increasing virtualities.

It is instructive to compare our approach with dispersive study based on the Roy-Steiner equations. In \cite{Hoferichter:2019nlq}, there is a different strategy for treating kinematic singularities and anomalous thresholds. Second, there is a coupling between $s$-wave and $d$-wave with strength related to the high-energy behavior assumption. Third, the extra subtraction in \cite{Hoferichter:2019nlq} leads to a $1/s$ singular behavior, which is due to the truncation of the p.w. expansion. In our approach, we solve a p.w. dispersion relation under the assumption of maximal analyticity. For the $s$-wave ($d$-wave), we perform a coupled-channel (single channel) dispersive analysis and present a simpler implementation of the anomalous thresholds. Furthermore, in this approach, there is no coupling between $s$- and $d$-waves and no extra $1/s$ singularities. In a work in preparation \cite{whitepaper:2019}, the comparison between \cite{Hoferichter:2019nlq} and our previous single virtual study \cite{Danilkin:2018qfn} together with current work has been done. Both approaches agree well up to the details due to a different treatment of the vector-meson couplings, form factors, and the inclusion of the coupled-channel in the $s$-wave.

\section{Conclusion}

In this work, we presented a dispersive analysis of the $\gamma^*\gamma^* \to \pi\pi$ reaction from the threshold up to 1.5 GeV in the $\pi\pi$ invariant mass. For the $s$-wave, we used a coupled-channel dispersive approach in order to simultaneously describe the scalar $f_0(500)$ and $f_0(980)$ resonances, while for the $d$-wave a single channel Omn\`es approach was adopted. The obtained results will serve as one of the relevant inputs to constrain the hadronic piece of the light-by-light scattering contribution to the muon's $a_\mu$~\cite{Colangelo:2017fiz,*Colangelo:2017qdm,*Colangelo:2014pva,Pauk:2014rfa}. Specifically it allows one to estimate the contributions from $f_0(500)$, $f_0(980)$, and $f_2(1270)$ resonances. The latter can be compared with the narrow resonance result \cite{Pauk:2014rta,Danilkin:2016hnh}. 

There are still a few open issues before it can implemented in a $(g-2)_\mu$ calculation. First, one needs to validate a current treatment of left-hand cuts by forthcoming BESIII data on the $\gamma \gamma^\ast \to \pi^+ \pi^-$ and 
$\gamma \gamma^\ast \to \pi^0 \pi^0$ reactions \cite{Redmer:2017fhg}. This is a prerequisite for a data-driven approach in quantifying the uncertainty of the HLbL  contribution to $a_\mu$. Second, for higher $Q^2$, one has to incorporate constraints from perturbative QCD for the vector transition from factors $f_{V,\pi}(Q^2)$ which is the driving force governing the $Q^2$ dependence of the $f_2(1270)$ resonance \cite{Danilkin:2019mhd}. This will be investigated in a future work.

\section*{Acknowledgements}
This work was supported by the Deutsche Forschungsgemeinschaft (DFG, German Research Foundation),
in part through the Collaborative Research Center [The Low-Energy Frontier of the Standard
Model, Projektnummer 204404729 - SFB 1044], and in part through the Cluster of Excellence
[Precision Physics, Fundamental Interactions, and Structure of Matter] (PRISMA+ EXC
2118/1) within the German Excellence Strategy (Project ID 39083149). O.D. acknowledges funding from DAAD.

\bibliographystyle{apsrevM}
\bibliography{PhysLettB_fast}

\end{document}